\documentclass[a4paper]{article}
 
\usepackage{amssymb}
\usepackage{amsmath}
\usepackage{latexsym}
\usepackage{minitoc}
\usepackage{hyperref}\usepackage{url}

\newcommand\fl[1]  {~\stackrel{#1}{\rightarrow}~}
\newcommand{\myset}[1] {\{ #1 \} }
\newcommand{\see}[1]{(see~ \ref{#1})}
\newcommand{\arxiv} [1] { \url{http://arxiv.org/abs/#1} }  
\newcommand{\arXiv} [1] { \url{http://arxiv.org/abs/#1} }  
\newcommand{\arXivs} [1] {\small  \url{http://arxiv.org/abs/#1} }  

\def\gr {general relativity}
\def\LCc {Levi-Civita connection}
\def\dmanif {differentiable  manifold}
\def\Weitzenc {Weitzenb\"ock connection}

\def \atit#1 {   \guill #1 \guillr} 
\def \btit#1 {   \emph{#1}} 
\def \booktitle#1 {   \emph{#1}} 
\def \rtit#1 {   \emph{#1}} 
\def\GL  {\textsf{GL}}
\def\SO  {\textsf{SO}}
\def\O  {\textsf{O}}

\def\SOf {\textsf{SO}^{\uparrow}}
\def\Vol {\mbox{Vol}}
\def\d {{\rm d}}
\def\Poinc{Poincar{\'e}}
\def\Poincg{Poincar{\'e} group}
\def\Poinca{Poincar{\'e} algebra}
\def\Minkv {Minkowski vector space}

\def\dS{de~Sitter}

\def\bydef{by definition}
\def\isom {~\stackrel{\sim}{\leftrightarrow}}

\def\fibre {\mapsto}
\def\fb {fiber bundle}
\def\mun  {^{-1}}
\def\dof {degree of freedom}
\def\dofs {degrees of freedom}
\def\calA {{\cal A}}
\def\T{{\rm T}}
\def\Diff {\mbox{Diff}}
\def\coord {coordinate}
\def\vf  {vector-field}
\def\munu { {\mu \nu } }

\def\guill {\textquotedblleft ~}
\def\guillr {\textquotedblright}

\def\calP {{\cal P}}
\def\Ad{\textsf{Ad}}  
\def\eqbydef {~\overset{\text{def}}{=}~}
\def\Aut{\textsf{ Aut}}  
\def\Rn{{\rm I\!R}^{n}}
\def\subgroup {\subset}
\def\Lieg {\underline{\mathfrak{g}}}
\def\Lieh {\underline{\mathfrak{h}}}
\def\wrt  {w.r.t.~}
\def\calC {{\cal C}}

\def\SL  {\textsf{SL}}
\def\Id{\mathbb{I}}
\def\calG {{\cal G}}

 \newcommand{\blit}[1] {{  \emph{  #1}}}

\def\spt {space-time}

\def\Fr {\mbox{Fr}}

\def\FrO {\mbox{Fr}^{\mbox{{\scriptsize    O}}}}
\def\FrSO {\mbox{Fr}^{\mbox{{\scriptsize     SO}}}}

\def\eg {\emph{e.g.}}
\def\ie {\emph{i.e.}}
\def\vsp  {vector space}

\addtocontents{toc}{\protect\thispagestyle{empty}}

\title{\bf \Huge  Connections and Frame Bundle Reductions    }
\author{M. Lachi{\`e}ze-Rey \\
 \small \it  APC-Astroparticule et Cosmologie (UMR-CNRS 7164),\\
\small \it   Universit\'e Paris Diderot-Paris 7,   10, rue Alice Domon et L\'eonie Duquet \\ \small \it  
F-75205 Paris Cedex 13, France\\
\small  mlr@apc.univ-paris7.fr
}
 \begin{document}

\maketitle
\abstract{
In \gr, the gravitational potential is represented by the \LCc,  the only symmetric connection  preserving  the metric. On a \dmanif,    a metric  identifies with  an orthogonal structure, defined as     a Lorentz  reduction of the frame  bundle. The \LCc ~appears as  the only symmetric connection  preserving  the reduction.
This paper presents  generalization of this process    to other   aproaches of  gravitation:  Weyl structure with  Weyl connections, teleparallel structures with     \Weitzenc s,   unimodular structure, similarly appear as   frame bundle  reductions, with preserving connections.  

To each subgroup $H$  of the linear group  $\GL$    correspond     reduced  structures, or H-structures. They  are subbundles  of the frame bundle (with $\GL$ as principal group), with $H$ as principal group.  A linear  connection in a manifold $M$  is  a principal connection on the frame bundle. Given a  reduction, the    corresponding \emph{preserving connections} on $M$ are the linear connections which preserve it.

I  also show that   the     \blit{time gauge}  used in the 3+1 formalism for  \gr ~similarly appears as   the result of a bundle  reduction.
 }\\


\section{Introduction}

Principal bundle reductions are   the mathematical expression of a physical process of symmetry reduction, from a principal group to one of its subgroups. They  play an important role in physics  and  are presented in many textbooks, mainly for  the archetypal case  of      symmetry breaking in gauge theories, a  variant  of the \emph{Higgs mechanism}. 

It is maybe not so well known   that  \gr ~may be seen as the result of  a similar  symmetry  reduction, from the    linear group to the   Lorenz group: a  metric on a manifold identifies with a reduction  from its  frame bundle to an  ON  tetrad bundle.  Then the     metric  connections are those (among the linear connections) which preserve this reduction.

This paper   shows how some  other approaches of   gravitation, with their relevant connections,  may be similarly considered  as resulting from a    reduction of the frame bundle.  This applies to Weyl theory  with Weyl connections; to teleparallel theories with   \Weitzenc s; to  unimodular theories.

In physics,  a  symmetry  corresponding to  a Lie group $G$  has a  \emph{local} expression  under the form of a $G$-principal bundle  $\calP$ (with   principal group~$G$), with the \spt ~manifold as basis. 
A  bundle reduction leads to  a $H$-reduced structure, with $H$  a subgroup of $G$. This is  a      $H$-principal subbundle, and a preserving connection is a  connection on $\calP$  which preserves the reduction.

The theory of \gr ~may be formulated for instance as a particular case of   orthogonal   reduction of the frame bundle $\Fr$, a $\GL$-principal  bundle 
  (see, \eg, \cite{Sardanashvily,SardanashvilyKurov}). A solution of the theory  is a Lorentzian    structure, \ie, a  $\SO$-principal  subbundle of $\Fr$, with the Lorentz group $\SO$ as principal group. This  identifies with the orthogonal tetrad bundle, itself equivalent to  the (Lorentzian)   metric. The preserving connections are the metric connections. Among them, the unique symmetric one, the \LCc, is identified to the gravitational potential.

The  section \ref{groupreduction}  first introduces  the \emph{group reductions}; in particular for a  linear group $\GL(E)$ acting on its representation \vsp ~$E$, as well as on the  space $B(E)$  of its vector  bases. A reduction of $\GL$     to a  subgroup~$H$   is defined as  a subspace of $B(E)$ which is an $H$-orbit. It may  then be  expressed  under the form of  some specific   structure  on $E$ (like for instance an inner product in the case of orthogonal reduction).

Section \ref{bundlereductions} recalls  the  general   process of   \emph{bundle reduction},  a generalization of group reduction. The section \ref{FrameBundle}   specializes    to the case of the frame bundle,    and presents the    generalizations (from global to local)    of the reductions presented in section \ref{groupreduction}. This make  appear different structures which have been proposed to express gravitation as results of such reductions: \gr, but also  Weyl theory, teleparallel  theory, unimodular theory...  The     \blit{time gauge}  in the 3+1 formalism for  \gr ~(which is used, for instance, in Loop quantum gravity) appears similarly as   the result of a bundle reduction.

Section \ref{Cartan} reminds  some remarks about the link between  Cartan connections and bundle reductions.

\section{Group Quotients and  Reductions}

\label{groupreduction}
\subsection{Group Quotients}

Throughout the paper, $G$ designs a  Lie--group (later we will specialize to  the  linear group $\GL$) of  dimension $D$ and  $  H$  a      subgroup  of dimension $d$, with  inclusion map $ H \fl{i} G$.  

We have a similar inclusion $\Lieh \to \Lieg$  of the corresponding Lie algebras; and $\Lieg$   may be splitten as  a  direct sum  of \vsp s,  $\Lieg =\Lieh \oplus   V$, with  $V$  a vector space (in general not an algebra) isomorphic to $\Lieg/\Lieh$. There are different ways to perform such splitting, and thus different isomorphisms~\footnote{ Since $\Lieh $ is a Lie algebra, we have 
 $[\Lieh , \Lieh]_{\Lieg}=[\Lieh , \Lieh]_{\Lieh} \subseteq  \Lieh .   $
 When in addition 
 $[\Lieh , V]_{\Lieg} = \Ad(H)\cdot V \subseteq  V  $,  then $V$ is $\Ad_H$-invariant 
 and the splitting is said reductive.}.

An  action of $G$ on some space  $M$ induces an   action of $H$ on the same  space. Any point $m\in M$  has an H-orbit $[m]  \eqbydef  \myset {h~m;~h\in H}$, which is an H-equivalence class in $M$. A  \emph{group-reduction}  is defined as  the choice   of  such  a class in $M$. The space $M$ itself is $G$-invariant but each class is only $H$-invariant: this is  a      symmetry reduction.

The set  of such classes (equivalently,  of  reductions) is the quotient  space  $B /H$,   
with   dimension $D-d$.  
 
\subsection{Linear Groups and Frame Reductions} 

We   specialize to the     linear group    $G=\GL(\Rn)=\Aut (\Rn)$, the group of linear transformations (automorphisms) of the \vsp ~$E=\Rn$, with dimension $D=n^2$. 

The linear group $\GL$ has also a natural action  on the set $ B(E)=\myset{b}$ of   vector  bases  (that I also call \emph{frames}) of $E$. Each  $g \in G$ identifies with a change of basis $  b\to g~ b$. This action is free and  transitive, so that    any choice of a preferred  basis (identified to the identity of $G$) provides   $B(E)$    with a group structure   isomorphic to   $\GL$  \footnote{ This means that $B(E)$ is a torsor, \ie, a group without unit element.}. 

A    subgroup $H\subgroup \GL$ also acts  on $B(E)$ and  associates to any  basis $b$ its   H-orbit $[b]  \eqbydef  \myset {h~b;~h\in H}$. Any choice of   such  an H-equivalence class $[b]$ in $B$  is  a symmetry reduction, from   the linear group $\GL$ to $H$.

The  set  of such classes, or equivalently the set of  reductions, is the quotient  space  $B /H$, with dimension $n^2-d$.  

In  section  \ref{FrameBundle}, such  group   reductions are extended to  frame  bundle reductions, by    performing them continuously    in each fiber of  the  frame bundle of a manifold. This   represents a  reduction of the corresponding   \emph{local}  symmetry. I  give before  some examples of frame-reductions.

\subsubsection{  Reduction to the orthogonal group}\label{orthogonalgroup}

 The typical example   is   $H=\O$, an orthogonal group \footnote{ I  do not specify now. This holds for    $\O(p,n-p)$; Later we will specialize to \spt, so that  $n=4$,    and to the Lorentz group $ \O(1,3)$. The procedure works for   any dimension and with any signature.}, with   dimension    $d=\frac{n^2-n}{2}$. The space   of possible reductions is   the quotient space $B/ \O$, with dimension $D-d=\frac{n^2+n}{2}\approx 10$ (I indicate with the sign $\approx$ the value for $n=4$, relevant for physical \spt).  

Each reduction  selects an  $\O$-equivalence class  $[b] $ in $B$:  a class of vector bases linked  together by   orthogonal transformations (which are the   elements of $\O$). 
That  class   defines     an unique inner product $p$ in $E$ (with the signature of~$\O$),  such that all its  frames    are   orthonormed (ON)  \wrt $p$. It results that  the $(D-d)$-dimensional set of  reductions identifies with      the set of   possible inner  products   in $E$ (see table   1).
 
 Note that an inner product  is defined by the components of the symmetric $n\times n$ matrix representing it (in a arbitrary basis). Their number is  $D-d=\frac{n^2+n}{2}$,   the dimension  of $G/O$.

\subsubsection{    Reduction to the Weyl  group}

We apply a similar procedure to   the \emph{Weyl  group}  $H=W\eqbydef  \Re ^+ \times \O$ (with $d=\frac{n^2-n}{2}+1 ~\approx 7$).
Here   $\Re ^+$ is  the multiplicative group; each element,  a positive real number, acts as   a  dilatation of the frame.  

Each reduction  corresponds to  a $W$-equivalence class $\ [b] \subset B$. Its   elements are  bases linked  together by  elements of $W$, \ie, orthogonal transformations as above, and dilatations. 

This defines not an inner product like above,   but a family $\calC $  of  inner products related by  multiplication by a positive real number. This family   exactly represents    a   \blit{ conformal structure} on $E$. The  frames of the class are those which   are ON      \wrt one of the inner products of  $\calC $.   The reduction is equivalent to this class, or to the conformal structure (see table  1).

A conformal structure   (a reduction) is defined by $D-d=\frac{n^2+n}{2}-1\approx 9$
numbers, the dimension  of $G/W$; one number less than for a metric,   corresponding to the dilatation freedom. 

\subsubsection{  Reduction to the identity}\label{trivial}

A  trivial reduction corresponds  to $H$ the group with  the  identity for unique element. The H-class of any basis   $b$  reduces to that basis  $b$ alone, so that   $B/H=B$. Each   reduction identifies with  the selection of   a single basis     $b_A \in  B$ (a choice of $n^2=     16$   numbers, the dimension of  $G/H=G$). Note that this   frame     defines   in turn an unique inner product \wrt which it is~ON.

\subsubsection{   Reduction to the special linear group}

Choosing the special linear  group $H = \SL$ ($d=n^2-1 \approx 15$), $\GL/H$ identifies with $\Re-\myset{0}$.  A reduction   corresponds to a  subspace $B^A \subset B$ of vector bases which share the same   determinant $A\ne 0$ (as expressed in  an arbitrary predetermined  frame).

Each reduction is expressed by  a  number $A$  and is  equivalent to the  class~$B^A$ of vector bases sharing the same determinant  $ A$ (see table  1). 
  
\subsubsection{   A double  Reduction gives  unimodular inner product}
\label{unimodularinner}

A  first  $\SL$-reduction, as in  the previous step,   defines the class~$ B ^A$ of the bases which share  the same  determinant $A$ (\wrt to a given fixed  basis).  

Then the  group inclusion $ \SO \fl{i} \SL$ defines the quotient space  $\SL/\SO$ with dimension  $\frac{n^2+n}{2}-1\approx 9$.  This allows   further reductions   by     selecting in  $ B^{A}$    a subclass $B^{AS}$ of bases    related by $\SO$ transformations (which  preserve the determinant $A$).

As above these bases define  an inner product $p_{AS}$. It   is however  constrained (by the first reduction) to have the determinant $\epsilon ~ A^2$ (in the predetermined  frame); the value of $\epsilon $ depends on the signature of $\SO$ ($\epsilon =1$ for Riemanian; $\epsilon=-1$ for Lorentzian). 
  
Finally,  the  double  reduction is equivalent to first choosing a number~$A$, and then  an inner product $p_{AS}$ having    determinant $\epsilon ~ A^2$. It is called \blit{unimodular} since the value of $A$ is usually taken to be 1  (in fact the choice of $A$ \wrt  a predetermined   reference frame is equivalent to the choice of a  reference  frame such that  $A^2=1$).

Extended to the frame bundle, this is at the basis of unimodular theories of gravity \see{unimodular}.

 \begin{table}[htp]\label{T1}
\caption{Reductions of the linear group $G=\GL$}
\begin{center}
\begin{tabular}{|c|c|c|c|c|}
\hline
subgroup & dim(H) &   dim(G/H)& reduction & reduced \\
H  &   $=d$& $=D-d$&  & frames\\
\hline
\hline
$\O $& $\frac{n^2-n}{2}$& $\frac{n^2+n}{2}$ & inner product   &ON \wrt  \\
 & &  & $  p$ &  \wrt $p$\\
\hline
\hline
$W $& $\frac{n^2-n}{2}+1$& $\frac{n^2+n}{2}-1$ & conformal structure    &ON \wrt    \\
 & &  & $[p]$ &   any $p\in [p]$\\
\hline
\hline
$\Id $& $0$& $ n^2 $ & one frame    &one  frame   \\
\hline
\hline
$\SL $& $n^2-1$& $1$ & $  A \in \Re-\{0\}$    &  frames with   \\
 & &  &   &  det. $= A $\\
\hline
\hline
\hline
$\SL  $& $ n^2-1$& $ 1$ & A    & fr. with det= $ A $\\
---&---&--- & ---    &---  \\
 $  \to \O $& $\frac{n^2-n}{2} $& $\frac{n^2+n}{2}-1$ &unimodular      &ON fr.  \wrt p\\
 & &   &  inner product    &  \\
 & &   &  with det $=\epsilon~A^2$    & with det= $ A $\\

\hline
\end{tabular}
\end{center}
\label{default}
\end{table}%


\section{Principal  bundle~reductions}
\label{bundlereductions}
A Lie group $G$ expresses a \emph{global} symmetry; a principal bundle  represents its \emph{local} counterpart. Its  automorphisms  form a group $\calG$ much larger than  $G$, often called the \emph{ local}  (not global)  symmetry (or gauge)  group. A  group reduction, as in previous  section,  may be generalized to  a  principal bundle reduction, which  corresponds to  a reduction of the local  (infinite) Lie group $\calG$. In particular   group reductions of $\GL$   generalize to reductions of the $\GL$-principal frame bundle.

The next section recalls the standard  construction of \blit{reduced structures}; then we  apply it  to the frame bundle of a \dmanif.  Hereafter  $M$ is a \dmanif ~of dimension $n$.

\subsection{Reduced Structures}

A $G$-principal   bundle  $\pi:~\calP  \fibre { M}$ has its   typical  fiber $F$    isomorphic to its principal group  $G$  (dimension $D $).  The    action    $y \to gy$ of $G$  on~$\calP$  defines that of its subgroup~$H$.   The latter  assigns to each $y\in \calP$  its   H-orbit  $[y]\eqbydef  \myset{ h y;~h \in H}$. This is an H-equivalence class in $\calP$ \footnote{ 
 Since the action of $G$, and thus of $H$, is vertical, the orbit $[y]$   is contained in $F_y$, the fibre over $y$.}
 and we have the natural projection $\pi _H:~ y \to [y]$.

 Physically, the $G$-principal   bundle $\calP$ represents a system with \blit{ local} symmetry $G$ and   $G$ is often called   the \blit{global} symmetry  group.  A reduction   is performed like above,   by     the choice of an   $H$-equivalence class \blit{in each fiber}. This choice will appear  equivalent to the construction of  a $H$-principal   bundle $\calP^ \sigma$. 

This represents  a local symmetry breaking from $G$ to $H$.  For instance, a choice of this kind is at the basis   of the Higgs mechanism in gauge theories, where  the reduced structure identifies with the Higgs field. 
Here we will  consider   different approaches to gravitation  as   similar reductions of  the frame bundle, from its original  local  $\GL$ symmetry to that of a subgroup.
This generalizes the well-known result that a Riemanian structure (a metric) identifies with an  orthogonal structure, \ie,   a reduction from the frame bundle to the   ON frame bundle (as we recall  below).

\subsubsection{General Construction of the Reduced Structure}

Each section   $s$ of $\calP  \fibre { M}$  
defines the map
$$m\to     \sigma(m) \eqbydef  (\pi _H \circ s)(m) \eqbydef   [s(m)] .$$ It  sends any basis point to 
the $H$-orbit of its image trough the  section~$s$. An important theorem  (see, \eg, \cite{Coq} p.119) states that 
$m\to   \sigma(m) $ is a section of the  \fb   
$$\Sigma \eqbydef     \pi _{\Sigma  }:~\calP/H \to M,$$ 
with $\pi =  \pi_H\circ    \pi _{\Sigma }  $. This  quotient bundle   has typical  fiber $F/H$  isomorphic to $G/H$  (dimension $D-d$).

Then each    section       $\sigma: ~M \mapsto \Sigma:~m\to  \sigma(m)   $
of this quotient bundle  defines a reduction, under the form of    the   subbundle of~$\calP$,
 $$\pi_\sigma:~\calP ^\sigma \to M  .$$   
This subbundle  is defined as  containing    only those elements    in $\calP$  which belong to the image of the section~$\sigma$:  
$$\calP ^\sigma      = \myset {y \in \calP, \exists m:~ y \in \sigma(m) } ,$$ with inclusion $\calP ^\sigma \fl{i}\calP$.    
Its  fiber  over~$m$ is  $(\pi_\sigma)\mun (m)\eqbydef  \sigma(m)$
 \footnote{ Locally, above an open $U_M$ of $M$, we have $\calP ^\sigma \simeq  U_M \times H$.}.
 
The  $H$-principal bundle  $\calP ^\sigma$ is the  [reduced]   \blit{$H$-structure}   associated to the section~$\sigma $. The construction shows the  one-to-one correspondence between  such reductions and     global sections~$\sigma$ of the quotient bundle $\Sigma  \eqbydef \calP/H \fibre  M$.

{\bf Degrees of Freedom}

One may count the \dofs: each  fiber of  $\calP$ has the  dimension  $D$  of   $G$: we have  $D$ \dofs ~for  chosing a section. 
The fiber  of  $\calP/H$ has  dimension   $D-d$: a section represents   $(D-d)$ \dof s. The fiber of  $\calP ^\sigma$ has   dimension   $d$: a section has $d$ \dofs.

In the language of gauge theories, a reduction  represents  a \blit{symmetry breaking}:~ from the  whole (un broken) symmetry group  $G$   to  the  group~$H$   of   broken  symmetries. A section   $\sigma$     can be interpreted as a classical Higgs field  \cite{Sardanashvily}.  
 Matter fields are sections of some vector bundle associated to~$\calP ^\sigma  $, whose fibers are    representations of $H$.
 
  \begin{table}[htp]
\caption{The \guill localization \guillr of Table 1: reductions of the Frame Bundle; \guill  frames \guillr ~hold for \guill moving frames \guillr }
\begin{center}
\begin{tabular}{|c|c|c|c|c|}
\hline
subgroup & structure&  reduction & reduced bundle& connection\\
H  &    &   &&   \\
\hline
\hline
$\O $& Orthogonal &   metric    &ON frames  & metric  \\
 &structure    &   g &  \wrt $g$&connection\\
\hline
\hline
$W $& conformal &    Conformal structure    &ON frames  & Weyl c.\\
 & structure &    [g] &  \wrt any $g\in [g]$&\\
\hline
\hline
$\Id $& Teleparallel&   one frame  $e$  & unique   frame $e$& Weitzenb\"ock      \\
&&&&connection\\  
\hline
\hline
$\SL $& &   $  A \in \Re-\{0\}$    &  frames with   &\\
  &  &   &  determinant $ A $&\\
\hline
\hline
\hline
 $\SL  \to \O $& Unimodular&  \guill unimodular \guillr       &ON frames   & connection\\
 & structure&     metric  $g$     & \wrt $g$& metric \\
 &&    with det $=\epsilon~A^2$    & with det= $ A $& \wrt $g$\\

\hline
\end{tabular}
\end{center}
\label{default}
\end{table}%

\subsection{Preserving  connections }

Now we assume a bundle reduction $\calP^H \fl{i} \calP$  linked to the choice of a section $\sigma$ of $\calP/H\to M$, as above.  Such a choice   fixes $d$ local (for each point)  \dofs ~ among $D$,   so that    $D-d$ remain.

On the other hand an  (Ehresmann)  principal   G-connection  on  $\calP$ is defined by a  connection form~$\omega$. This is a one-form in $\calP$ 
\footnote{ It does not depend on a trivialization, in contrary to  \emph{ local} connection forms, which are defined  in the basis manifold $M$.} with values in~$ \Lieg$. We may demand  that   the connection preserves the reduced structure. This  is equivalent to demanding  that     $\omega$ takes its value  in $\Lieh \subset \Lieg$ only. We refer to this case as   a \blit{preserving connection}.  We show below how specific   types   of linear connections   in manifolds, in relation to some gravitation theories,  identify with particular preserving connections. 

The space $\calA$ of   (smooth)    connections in $\calP$  admits    $n~ D\approx 64$ local  \dofs.
For a given  reduction $\sigma$, the space $\calA _\sigma$ of preserving connections (also in   $\calP$) has only    $n~d$  local \dofs ~  
\footnote{ This may be seen from the fact that the  reduced structure corresponds to  $D-d$ \dofs. The preservation conditions represent $n~(D-d)$ equations in the $(n~D)$-dimensional space of connections, leaving a space of preserving connections with dimension $n~d$.}.

A preserving connection is  a particular case of  \emph{reductive connection}. For the later, the connection form admits an  $\Ad_H$-invariant splitting 
$\omega = \omega ^h + \omega ^v$, where $ \omega ^h$   takes its  values in $\Lieh$.
This implies that $i^* \omega ^h $  is a principal H-connection in~the H-principal bundle $\calP^H$:  it takes its  values in   $\Lieh$ and is H-equivariant. The case of a reducing connection here corresponds to~$\omega ^v=0$.

We  apply below  to  linear  connections in a manifold, which are   in fact  $\GL$-principal connections  in the  frame bundle.

 \section{Reductions of the  Frame Bundle}\label{FrameBundle}
\subsection{The Frame Bundle}

The   frame bundle  $  \Fr (M)\fibre M$ of   a manifold $M$  is a    GL-principal \fb ~with  $G=\GL$ the linear group, and $D=n^2$.   Now I write $\Fr$ for $\Fr(M)$.  Its  fiber over  $m$ is the set   $\Fr_m \eqbydef \myset{f_m} $ of   possible vector bases    of $\T_mM$, the tangent space at~$m$; $\Fr_m$  is (non canonically)  isomorphic to $\GL$ and we recover in each fiber the situation of the first section.

A section of $\Fr$   is  a \emph{moving frame}   $f:~m\to f_m \in \Fr_m$. It  assigns to each $m $ a basis $f_m$  of $\T_m M$.    A smooth \emph{local} action of $G=\GL$ (the action of an element of $\GL$ at each point of $M$) identifies with  a change of moving frame. It is free and transitive.   Such local actions  form the  infinite-dimensional group $\calG$ \footnote{
 It admits [the pullback of]  the infinite Lie group $\Diff$ as a subgroup 
Note that the reduction of linear moving frames to \emph{holonomic} frames (or \coord ~frames) may be seen as a reduction of $\calG$ to $\Diff$.}.

\subsubsection{Linear  Connections }

A linear connection on a manifold $M$  is a  principal connection on the frame bundle,  with $G=\GL$.  The space $\calA$  of [smooth] linear connections has  $n~D=n^3\approx 64$ local \dofs.

One may  describe a  linear connection   by its local connection form (in a given fixed  moving frame),  a \GL -equivariant one-form $\omega$ taking its value in $\Lieg$. Since $G$ is the linear group, the components of $\omega$  are usually  written with manifold  indices (in the same frame)  as 
$\Gamma ^\alpha_{~\beta}=\Gamma ^\alpha_{\mu\beta}~ \theta ^\mu$. The  connection is usually  represented by its connection coefficients $\Gamma ^\alpha_{\mu\beta}$. Its  3 manifolds indices  represent the $n^3$ \dofs ~of~$\calA$.

\subsection{Orthogonal reduction}

The well known case (see, \eg, \cite{Sardanashvily,SardanashvilyKurov}) of \blit{ orthogonal reduction} corresponds to the  choice  of an orthogonal group $H=\O$, with $d=n~(n-1)/2 $ \dofs ~(the procedure works for   any dimension and with any signature).   Each  reduction        defines an orthogonal structure.

We apply a   reduction  like in \ref{orthogonalgroup}, but  fiberwise. A fiber of $\Fr/ \O \to M$  is the  quotient  $  \Fr_m/ \O$, the set  of O-equivalences classes  of local  bases; or equivalently, of $\O$-orbits in $\Fr_m$.  A section $\sigma$  of $\Fr/ \O \to M$  is  a continuous choice   $\sigma(m)$ of  such a class for each $m$. Like above,  all the  frames  in  the class $\sigma(m)$ are  linked together  by  O-transformations, and   they    appear as  ON    \wrt an  inner product  $p_m$ in  $\T _mM$.    The reunion of these inner products (one for each $m$)    builds a metric $g$ over~$M$.  Finally, the reduction, and equivalently  the     section of $\Fr / \O$,   identifies with  
 the metric $g$. For this reason,  $\Fr / \O$ is called   the \blit{metric bundle}. We have the equivalences \\
{\bf reduction  $\simeq$       section of $\Fr/ \O$   $\simeq$ metric  $g$ $\simeq$ class   of ON moving frames (tetrads)  \wrt $g$.}

Each   reduction (choice of $g$) defines the related \blit{ orthogonal structure}  $\FrO(M) \subset \Fr(M)$,  called the       {tetrad  bundle}, or ON ~frame bundle,  corresponding to $g$. This     sub-bundle of $\Fr$  contains, as sections,  only  the  moving frames taken in the class, \ie,  ON tetrads
  \wrt ~$g$.   It admits the structure group  $ \O$.    
  
 {\bf Metric  connections }

Preserving connections  preserve  the   orthogonal structure: they are  the   \emph{metric connections}. Given a  metric, the  space of metric   connections   admits    $ n^2 ~ (n-1)/2  \approx  24$ local \dofs. 
\footnote{ Preserving the 
$D-d$ \dofs ~of the metric correspond to 
$n\times  (D-d)  $ equations 
 that the connection must obey.  Since the space of connections has  $n^3$ local \dofs, this leaves $d~n  $  a space of metric connections.}
 
 Asking  in addition  for  \emph{symmetry} (no torsion)  fixes   all the  remaining \dofs, resulting in the  unique \LCc ~for that metric.     \footnote{ The space of symmetric connections has   $n~(n^2+n)/2 \approx 40 $ local \dofs.} 

  \subsubsection{ Lorentz Structure and Time Gauge}
\label{timegauge}

The case  $n=4$ and the choice of   the Lorentz group $\SO(1,3)$  corresponds to \spt ~and \gr. A   \blit{Lorentz structure} is  an other name for  a Lorentzian metric $g$, with    $\FrSO$  identified with  the corresponding  ON tetrad bundle
\footnote{
 For any signature,  the  group  inclusions  $\SOf \subgroup \SO\subgroup \O\subgroup \GL $
allows us to  replace $\O$ (reduced O structure $\FrO $ of ON frames)  by  $\SO$ or  $\SOf$. One   obtains  respectively:\\
- the  reduced $\SO $ structure $\FrSO $ of     oriented ON frames;\\
- the   reduced $\SOf$  structure of      oriented  and time-oriented ON frames. }
on \spt, with  $D=\frac{n^2-n}{2}\approx6$ \dofs. Each section is a tetrad.
\guill  The configuration space1 of general relativity can thus be seen as the space of sections of the bundle $\Fr / \SO$    \guillr \cite{JANSSENS}.

  We may extend to a new  reduction to   $\SO(3)\subgroup \SO(1,3)$. This defines an  $\SO(3) $-structure  on \spt, under the form of    an $\SO(3)$-principal subbundle  of $\FrSO $.

We have  $d=\frac{(n-1)^2-(n-1)}{2} \approx 3$.  The space of such reductions is  the quotient bundle  $\FrSO/\SO(3)$, with  dimension $D-d\approx3$ per fibre. This is the number of \dofs ~of the space of reductions.

A section of the reduced structure  is a class $\Fr^u$ of  ON tetrads  which are     linked  together by  the (local) orthogonal transformations of $\O(3)$, \ie,   the spatial rotations in each fiber.   In other words all  the tetrads  of this class  share the same  timelike unit \vf~$u$.  And the   reduction identifies with the   choice of this  \vf ~u. This process is called   a \blit{time gauge} (see also \cite{Fatibene}).   The $\SO(3)$-structure is the corresponding   \blit{triad bundle} \footnote{ Each  tetrad is decomposed as $u$ plus the (spacelike)  triad.}.

 \subsection{Weyl group and Weyl geometry}

The choice of  the Weyl group     $H=W$ (with  $d=n~(n-1)/2+1\approx 7$) defines  the  reduction  to a Weyl structure, or conformal  structure. This is   a class $[g]$ of metrics related by [Weyl] scalings (\ie,   multiplications by a  scalar positive function).  Any   moving frame  in the reduced structure is ON \wrt one of the metrics of the class $[g]$.
Such a \blit{ conformal structure} represents  $\frac{n^2+n}{2}-1 \approx 9 $ \dofs. 

A connection   preserving   the  conformal structure $[g]$ is called  a  \blit{Weyl connection}.
For a given reduction, the  space of   {Weyl connections} admits      $ n^2 ~ (n-1)/2+n \approx  28$ local \dofs. Imposing symmetry (no  torsion), as usual, fixes $ n^2 ~ (n-1)/2 \approx    24$ \dofs ~among them. The $n\approx   4$ remaining   correspond to the choice     of a \blit{Weyl form} $A$ such that, for any member $g$ of the class $[g]$,    $\nabla   g = A \otimes  g $  (the form $A$ depends on the representative of the class). This corresponds exactly  to a \blit{Weyl geometry}. It  is  \emph{integrable} when  $A$ is an exact  form.

\subsection{Unimodular Theory }\label{unimodular}

The frame bundle is a principal $\GL$-bundle. The  reduction of $\GL$ to  $\SL$ (generalizing \ref{unimodularinner}) fixes  1  \dof . It    is well expressed  by  the choice of a volume form $\Vol$
\footnote{ 
This is  equivalent  to the  choice of  a scalar function in an arbitrarily  chosen moving frame; but the formulation with  a volume form offers the advantage of  remaining covariant. Also, it is always possible to chose the  reference  frame such that $f$ is a constant function equal to unity; hence the appellation \emph{unimodular}
\cite{Henneaux}.}.  A section of the $\SL$ structure is  the class  of moving frames whose associated volume    element      identifies with~$\Vol$.    Such frames are linked together  by the local \SL ~group.

Then a subsequent reduction of  $\SL$ to an orthogonal group $\SO$  (like  the  Lorentz group) 
fixes $n~(n-1)/2 \approx 6$ additional local  \dofs. A     section is  a family of      moving frames linked by orthogonal transformations and sharing the  same volume form   $\Vol$.  Like in \ref{unimodularinner}  this defines    a    metric~$g$, which is   constrained to obey $\Vol_g=\Vol$.  The space of such unimodular metrics    admits   $n~(n+1)/2-1 \approx 9$  local   \dofs. 

A metric  connection (like the \LCc) \wrt  an unimodular metric  preserves both structures.

\subsubsection{  \Weitzenc s}

A    trivial reductions of $\GL$ to to identity  
 --- a generalization of   \ref{trivial} --- selects an unique basis $f_m$ in  each fiber $\Fr_m$; thus a moving frame $f$  corresponding   to  a section of $\Fr=\Fr/ \Id$.  Note that such a   moving frame, and thus the reduction,  only exists (when  continuity is required) for a  parallelizable manifold. 
  The reduced structure  identifies with that moving frame, sometimes  called       the AP-frame, for \guill Absolute Parallelism~ \guillr.  In other words, the reduction   fixes the totality of the   \dofs~in~$\Fr$~\footnote{ Preserving the  16 \dofs ~of a tetrad correspond to $ 4 \times 16=64$ equations; which fix the 64 \dofs ~of a connection.} and  breaks totally  the linear symmetry. 

It turns out that there is an   unique \footnote{ In dimension 4,  preserving the 16 \dofs ~of a moving frame  correspond to $ 4 \times 16=64$ equations, which fix   the 64 \dofs ~of a the possible connections, thus leading to the same conclusion.}  connection preserving the AP-frame (\ie, the reduced structure): $\nabla e=0$, which   means  $\nabla (e_I)=0,~\forall I$. It is  called the  \blit{\Weitzenc}~associated to the selected AP-frame~$e$.
This  implies    zero  connection coefficients      in the AP-frame, from which    follows ---
as it is well known ---  that  this connection is flat (with zero   curvature)\footnote{  Flat connection  do not exist in any  \dmanif.}.  This connection has however    torsion,  described  by the  local torsion form $T^I=\d e ^I$ in the AP-frame. 

Like any coframe, the reciprocal coframe $\theta$   of the AP-frame  defines in turn    an  unique metric $g_\munu \eqbydef \eta_{IJ}~ \theta^I~ \theta^J$,  \wrt which it is  ON.
This  AP-metric  is also  preserved by the  \Weitzenc ~which is thus  metric  \wrt  $g$.  
Of course, the later  differs from the  \LCc ~ associated to $g$ but, as it has been emphasized   a long time ago by Einstein himself,     it is possible to  construct from  the torsion of the \Weitzenc ~  an action which leads to  equations for the metric similar  to that of \gr ~(obtained from the  \emph{Einstein-Hilbert action} constructed with  the curvature scalar of the \LCc).  This  analogy between this approach, named   \emph{teleparallel}, and \gr ~may however  hold   only  when a \Weitzenc ~exists, and  thus for parallelizable manifolds. 

The  main difference is that, by construction,     the teleparallel view   breaks explicitely the   Lorentz symmetry through  the choice of  the AP-tetrad:  it breaks  totally  (not partially) the  linear symmetry, although \gr ~breaks it    to Lorentz  symmetry only~\footnote{ 
One may obtain a teleparallel     structure, with associated  \Weitzenc,   from  a double reduction: first an orthogonal reduction, from the linear group to the orthogonal (Lorentz) group; then a second reduction entirely breaking the remaining   orthogonal (Lorentz)  symmetry, to preserve the AP-frame.}. 
Thus, a  physical equivalence can only associate  teleparallelism with   \gr ~plus  an additional physical  entity corresponding to the breaking of   the Lorentz symmetry.

A link has also  been  suggested 
with a gauging of  translations (see \cite{PereiraObukhov, Huguet} and references therein).  A moving coframe $\theta$  is a one form taking its value  in $\Re^4$.   The identification of the latter  with  the Lie algebra  of the group of translations \cite{Baez} may allow   to  consider $\theta$ as a principal connection for      this group.  This   is  however        not a \emph{linear} connection on the manifold and, in particular, torsion  is not defined for it. It turns out however  that the   curvature form, $\d \theta$,   of this connection identifies with the torsion form of the \Weitzenc ~expressed in   the same tetrad (considered as AP-tetrad).  This mathematical analogy lacks however  geometrical  support since  the involved  \guill translations \guillr ~ have  absolutely no geometrical action,  on  \spt ~or on  its tangent space. This suggest that one should rather refer to  an   \emph{internal $\Re^4$ gauging}   than a  \emph{ translation gauging}.

\section{Some Remarks about Cartan Connections }
\label{Cartan}

Let me finish with  some remarks concerning  \emph{Cartan Connections}, which are also linked to bundle reductions, although in a different way. 

I assume a bundle reduction $\calP^H\fl{i} \calP $, with $\calP$  a G-principal bundle and $\calP^H$ a H-principal bundle.
To fix the ideas,    $H$ may  be an orthogonal group $\O$ and  $\calP^H$    the orthogonal frame bundle like the usual \emph{tetrad bundle}  in Riemanian geometry; and $G$   a larger group, including $\O$ as a subgroup,  typically \Poinc, \dS ~or anti-\dS. Thus the resulting geometry is often called the gauge theory of these groups \cite{WISE}.

 A principal connection form $\widetilde{ \omega} $ on $\calP$
footnote{
We  precise that the G-principal bundle $\calP$ is  only  used as an auxiliary for the derivations and has not necessarily a physical interpretation:  the physics is described in the H-principal bundle $\calP ^\sigma$. The derivation of the Cartan geometry may be  entirely formulated in it  without   implying the bundle $\calP$. It involves a $\Lieg$-valued connection (not a principal connection)  $\omega$,  which takes its  values in $\Lieg$ (not in $\Lieh$).  
} takes  values in~$\Lieg$. Like above we split $\widetilde{ \omega} =\widetilde{ \omega} ^H+\widetilde{ \omega} ^V$,
with $ \widetilde{ \omega} ^V $ taking its values in   a vector space   $V$ (in general not an algebra) isomorphic to $\Lieg/\Lieh$
\footnote{ Since $\Lieh $ is a Lie algebra, we have 
 $[\Lieh , \Lieh]_{\Lieg}=[\Lieh , \Lieh]_{\Lieh} \subseteq  \Lieh .   $
 When in addition 
 $[\Lieh , V]_{\Lieg} = \Ad(H)\cdot V \subseteq  V  $,  then $V$ is $\Ad_H$-invariant 
 and the splitting is said reductive.}. 

Then  $\omega\eqbydef i_* \widetilde{ \omega}$  is \bydef ~a Cartan connection in $\calP^H$
\footnote{ Some authors  call $  \omega$   a Cartan connection on $\calP$.} 
iff it verifies the   \emph{Cartan condition}:\\
\indent {\bf for all $p\in \calP^H$, $ \omega $   defines a linear isomorphism $\T _p \calP^H \isom \Lieg$}.\\ This requires $D=d+n$. 
Taking its  values in $\Lieg$, $\omega$  cannot be  a principal connection  since the principal group of $\calP^H$  is $H$. It is however H-equivariant \cite{CATREN}.
\footnote{ Note that a    Cartan connection $ \omega $ can also be defined intrinsically  in a H-principal bundle  $\calP^H $  as\\
-  taking  values in $\Lieg$ and being H-equivariant;\\
- verifying the Cartan condition above;\\
- and giving the value $ \omega (\zeta)=\underline{\zeta}$ for the fundamental \vf ~$\zeta$ (Killing \vf) in $\calP^H$  corresponding to $\underline{\zeta}\in \Lieh$.}

When in addition  the splitting $ \omega = \omega ^H+ \omega ^V$ is   \emph{reductive} (\ie,  also  H-equivariant), then  $ \omega ^H \eqbydef  \i ^* \widetilde{ \omega} ^H$ is a H-principal  connection  on $\calP^H$, called a \emph{reduced connection} and  $e \eqbydef    \i ^* \omega ^V$ a H-value one form on $\calP^H$ called the \emph{solder form}.

\subsection{Gauging of the  \Poinc ~and \dS ~Groups }

This applies in particular to  $G$ the \Poincg,  and   $H=\SO(1,3)$ the Lorentz group. We assume $n=4$.
 
$\calP $ is a principal $G$-bundle over \spt ~$M$. The reduced  Lorentz  structure is  $\FrO$, the tetrad bundle  over $M$,   equivalent to a Lorentzian manifold with metric $g$.  A principal connection $\widetilde{\omega}$ over $\calP $ takes its  values in $\Lieg$. It  does not generate a linear connection over $M$ since this is not a linear connection on its frame bundle.

The commutator of a translation with a rotation being  a translation,  the \Poinca ~admits a reductive splitting  $\Lieg =\Lieh +V$. 
The connection $\widetilde{\omega}$  splits as  $\widetilde{\omega}=\widetilde{\omega}^H+\widetilde{\omega}^V$, with the isomorphisms 
$\T_m M \isom \widetilde{\omega}^V \isom \Lieg\Lieh$, thus obeying the Cartan condition.

We obtain the  Lorentz-invariant splitting   $i_*\widetilde{\omega}\eqbydef \omega= \omega ^{SO}+{\omega}^{V}$. Here $ \omega$ is the Cartan connection on $M$ (not a linear connection),    and    the  principal   Lorentz  connection
$\omega ^{SO}$ defines a linera metric connection on~$M$. The  \Minkv-valued one-form     ${\omega}^{V} $ in $\FrO$ is the solder form,  the bundle expression of the cotetrad. 
Note that  the present construction also holds with the \Poincg ~replaced by \dS ~or anti-\dS ~groups.

  
\end{document}